# Toward a Calculus of Redundancy:
# Signification, Codification, and Anticipation in Cultural Evolution

Loet Leydesdorff,[a]* Mark W. Johnson,[b] and Inga Ivanova[c]

**Abstract**

This paper considers the relationships among meaning generation, selection, and the dynamics of discourse from a variety of perspectives ranging from information theory and biology to sociology. Following Husserl's idea of a horizon of meaning in intersubjective communication, we propose a way in which, using Shannon's equations, the generation and selection of meanings from a horizon of possibilities can be considered probabilistically. The information-theoretical dynamics we articulate considers a process of meaning generation within cultural evolution: information is imbued with meaning, and through this process, the number of options for the selection of meaning in discourse proliferates. The redundancy of possible meanings contributes to a codification of expectations within the discourse. Unlike hard-wired DNA, the codes of non-biological systems can co-evolve with the variations. Spanning horizons of meaning, the codes structure the communications as selection environments that shape discourses. Discursive knowledge can be considered as meta-coded communication which enables us to translate among differently coded communications. The dynamics of discursive knowledge production can thus infuse the historical dynamics with a cultural evolution by adding options, that is, by increasing redundancy. A calculus of redundancy is presented as an indicator whereby these dynamics of discourse and meaning may be explored empirically.

[a] * corresponding author; University of Amsterdam, Amsterdam School of Communication Research (ASCoR), PO Box 15793, 1001 NG Amsterdam, The Netherlands; email: loet@leydesdorff.net; ORCID: 0000-0002-7835-3098.
[b] University of Liverpool, Centre for Educational Development and Support, Faculty of Health and Life Sciences, Brownlow Hill, Liverpool UK; email: mwj1@liverpool.ac.uk;
[c] Institute for Statistical Studies and Economics of Knowledge, National Research University Higher School of Economics (NRU HSE), 20 Myasnitskaya St., Moscow, 101000, Russia; and School of Economics and Management, Far Eastern Federal University, 8, Sukhanova St., Vladivostok 690990, Russia; inga.iva@mail.ru ; ORCID: 0000-0002-5441-5231.

**Introduction**

In this study, we analyze the relations between information processing and meaning processing in interhuman communications. Different meanings can be provided to Shannon-type information. In interhuman communications, however, meanings are provided with reference to intersubjective horizons of meaning spanned by codes in the communication. Our objective is to specify and operationalize the relations among these three dynamics of (1) information processing, (2) the processing of meaning, and (3) the codification of communication. When symbolically generalized, the codes can be expected to operate as coordination and selection mechanisms—reducing uncertainty—whereas information processing generates uncertainty. The operationalization in terms of mechanisms potentially generating redundancy will enable us to specify an indicator for the trade-off between the evolutionary self-organization of codes in inter-human communications and their historical variation in instantiations.

Although the analysis has philosophical overtones, we do not intend to formulate an encompassing theory. We bring together specific insights from various perspectives relevant for developing a calculus of redundancy: information theory, biology, evolution theory, the theory and computation of anticipatory systems, and sociology. Can a calculus of redundancy be envisaged, complementary to the calculus of information provided by Shannon's (1948) mathematical theory of communication (Bar-Hillel, 1955; Theil, 1972)? In our opinion, the analytical distinctions between information, meaning, and knowledge as codified meaning can be used to enrich the theoretical framework and can be operationalized.

The meaning of information can be changed independently of the information content of a message, and the information content can change without consequences for its meaning. In other words, a degree of freedom is added to the communication when one assigns meaning to a Shannon-type information $H$—formalized as $H = -\sum_i p_i \log(p_i)$, where $p_i$ is the relative frequency distribution of a random variable.[1] The maximum entropy of systems which process both information and meaning is not $\log(N)$—the logarithm of the number of options $N$—but $\log(M * N)$, where $M = \sum_{i=1}^{N} m_i$, the sum total of possible meanings provided to an information $i$.[2] Independently of the development of the information, the maximum entropy is thus enlarged by increasing the number of options. In other words, redundancy is not a given, but is constructed by defining and codifying relevant categories. Consequently, the processing of meaning in inter-human communication can be expected to enlarge (or reduce) the number of options.

---

[1] If two is used as the base for the logarithm, the information is expressed in bits of information.

Redundancy was defined by Shannon as the complement of uncertainty relative to the maximum information content of a system. In electrical engineering, redundancy is needed for error correction. In common language, however, "redundancy" connotes superfluous and repetitive messaging. Redundancy can also be considered as "excess information" in a message: the message could have been shorter if less redundant (Shannon, 1945). Shannon's co-author Weaver (1949), however, noted that redundancy might be a prime candidate for the development of a theory of meaning, since what is considered as error assumes a perspective—that is, another discourse—from which the information can be provided with meaning ("signal") or discarded as noise.

Bateson (1972, p. 420) argued that "the concept 'redundancy' is at least a partial synonym of 'meaning:' […] if the receiver can guess at missing parts of the message, then those parts must, in fact, carry a *meaning* which refers to the missing part and is information about these parts. […] In sum, 'redundancy' and 'meaning' become synonymous whenever both words are applied to the same universe of discourse." Whereas this "same universe of discourse" is species-specific (Maturana, 1978), interhuman communications provide meanings not only reflexively but also with reference to potentially different "horizons of meaning" (Husserl, 1929). This further codification in potentially different directions enriches our notion of meaning in interpersonal relations with intentionality; that is, beyond Bateson's definition of meaning as synonymous to redundancy in a single universe of discourse. We shall argue in this study that meaning in this case is not only a local "incursion" (Dubois, 1998) of redundancy, but at the same time an operation that mediates the trade-off of uncertainty generation with the arrow of time ("recursion") and the feedback of codes of communication against the arrow of time ("hyper-incursion"). This trade-off can be measured in bits of information.

---

[2] $\sum_{i=1}^{N} m_i$ is real valued function ($m \geq 1$).

The *dynamics* of information processing and meaning processing are also different. Whereas information processing is grounded in the second law of thermodynamics, the dynamics of meaning are discursive: meanings are constructed reflexively. Against the monist programs and philosophies nowadays prevalent in artificial intelligence (e.g., Damasio, 2006), we thus argue in favor of a dualism between information and meaning. Although both can be considered as constructs, their status is very different. Unlike information, meaning is not communicated. (Were meaning communicated, it could again be considered as information.) Communications are relational; providing meaning presumes a position from which a perspective on the communications can be entertained (Burt, 1982). Meanings can be shared to differing extents when perspectives overlap. The sharing of meanings can be expected to provide a source of redundancy.

**The perspective of information theory**

Using information theory, the maximum information content of a system ($H_{max}$) is determined by the total number of its possible states $N$; that is, $H_{max} = log(N)$. $H_{max}$ is composed of two parts: the states of a system hitherto realized ($H_{system}$) and the states that are possible given the definition of a system but have not yet been realized. This surplus is redundant. Redundancy $R$ is formalized as follows:

$$R = \frac{H_{max} - H_{system}}{H_{max}} = 1 - \frac{H_{system}}{H_{max}} \qquad (1)$$

While the information content and redundancy are thus anchored in information theory as complements, the relation with the intuitive concept of redundancy as repetition requires a translation because redundancy is then conceptualized as a *dynamics* different from information processing.

Using the *H* in Gibb's formulation of entropy ($S = k_B * H$),[3] Shannon (1948) chose to define information as probabilistic entropy. As a consequence, the development of information follows the second law of thermodynamics and can therefore only be positive (Krippendorff, 2009a). The generation of redundancy, however, can result in a positive or negative contribution to the flow of entropy depending on the feedback and feedforward loops in the meaning processing as different from the information processing (Krippendorff, 2009b). Feedback and feedforward loops can propel information and meaning in clockwise or counter-clockwise cycles; that is, with potentially opposite signs (Ulanowicz, 2009; Ivanova & Leydesdorff, 2014). The relative information content of a message ($H_{system}/H_{max}$) can thus be enlarged or reduced. In sum, options other than those already historically realized can be added to or removed from a system by mechanisms very different from the second law.

Shannon (1948) distanced himself from the study of meaning-processing in feedback loops by declaring that "[t]hese semantic aspects of communication are irrelevant to the engineering problem." Weaver, however, speculated that Shannon's analysis "has so penetratingly cleared the air that one is now, perhaps for the first time, ready for a theory of meaning" (Shannon & Weaver, 1949, p. 27); but he added emphatically that "(i)nformation must not be confused with meaning" (Shannon & Weaver, 1949, p. 8). Over the half century since then, however, confusion has prevailed about the definitions of information, meaning, and their relation. Meaning has mainly been discussed in relation to information and not in terms of redundancy (Hayles, 1990, pp. 59 ff.). Meaning is provided intentionally and reflexively with reference to horizons of meaning. These horizons can be considered as intentionality at the inter-subjective or supra-individual level (Husserl, 1929).

The generation of redundancy is not observable naturalistically and can therefore not be the subject of positive theorizing. States other than the historically observable ones can be added reflexively as empty boxes—in other words, observational categories still to be filled out. The number of options in a social system can thus increase much faster than their realizations: one can assume both of the roads "diverged in a yellow wood" (Robert Frost). Entertaining maps of these roads and the simulation of possible occurrences changes the meaning of the roads that were actually taken. In a model, realizations are considered as instantiations among ranges of possible states. Models enable us to envisage and construct new technologies; new technologies can be expected to add not only specific options (variants of existing technologies) but mew degrees of freedom to the development, and thus to multiply the number of options.

---

[3] $k_B$ is the Boltzmann constant which provides the thermodynamic entropy *S* with the dimensionality Joule/Kelvin. The Shannon entropy $H [= -\sum_i p_i \log(p_i)$ ; see above] is dimensionless.

As long as transportation over the Alps, for example, is constrained by passes such as the Brenner and the Gotthard, the number of these passes determines the maximum capacity. Railways which can be tunneled under the mountains and airplanes which can cross the Alps independently of conditions on the ground multiply the number of options. These options, however, have to be invented and their realization requires a plan. A plan anticipates on future states; furthermore, a plan can be communicated and improved. Meanings can be coded differently in the various theories needed for constructing the models. Translations among differently coded communications can be expected to generate both redundancy and uncertainty. As argued above, providing meaning adds options to the system. However, this proliferation is constrained by codifications. Meanings tend to proliferate; codes of communication can be expected to reduce the prevailing uncertainty as selection mechanisms.

**"A difference which makes a difference"**

Shannon's counter-intuitive definition of information as uncertainty induced a call for an intuitive definition of information as "reduction of uncertainty" or neg-entropy (Schrödinger, 1944; Brillouin, 1962). Varela (1979, at p. 266) argued for defining "information" in accordance with the semantic root of the word of "in-formare." In a similar vein, the anthropologist Bateson proposed defining information as "a difference which makes a difference" (Bateson, 1972, at p. 315; cf. MacKay, 1969). However, a difference may make a difference for one system of reference but not, or differently, for another. In other words, information would then no longer be defined analytically, but in terms of what information means for a receiving system. In second-order systems theory, this receiving system has been denoted as an "observer" (e.g., Maturana, 1978; Von Foerster, 1982).

In sum, a "difference which makes a difference" defines "meaningful information," where the receiving system is the system of reference. Receiving systems can be studied empirically in terms of behavior. In our opinion, the communications themselves and not the communicators (or the observers) should be considered as the units of analysis. When the sending and receiving systems are considered as the systems of reference and information is defined as "a difference which makes a difference" for them, the definitions of information, meaning, and discursive knowledge become entangled. The more abstract perspective of information theory on the number of options and the measurement of realized ones in terms of (e.g., bits of) information tends to be lost.

In a recent book, César Hidalgo (2015, at p. 165), for example, has defined "information" with reference "to the order embodied in codified sequences, such as those found in music or DNA, while *knowledge and knowhow* refer to the ability of a system to process information." However, codified knowledge can be abstract and—like music—does not have to be "embodied" (e.g., Cowan, David, & Foray, 2000). Beyond Hidalgo's position, the philosopher Luciano Floridi (2010, p. 21) proposed "a general definition of information" according to which "the well-formed data are *meaningful*" (italics of the author). The sociologist Niklas Luhmann (1995, p. 67) posits that "all information has meaning." In his opinion, information should therefore be considered as a selection mechanism. Kauffman *et al*. (2008, at p. 28) added to the confusion by defining information as "natural selection."

Against these attempt to bring information and meaning under a single denominator, we argue for a dualistic perspective. We elaborate on Deacon's (2012, at p. 3) call for a focus on "the absent": "What is absent matters, and yet our current understanding of the physical universe suggests that it should not. A causal role for absence seems to be absent from the natural sciences." Deacon added (at p. 9): "This something-not-there permeates and organizes what is physically present in these phenomena. Its absent mode of existence, so to speak, is at most only a potentiality, a placeholder. […] Zero is the paradigm example of such a placeholder." Beyond this position, we argue that the social sciences not only fill in the gaps (the zeros) by specifying expected values, but can provide us with insight into the potentially negative values of the intangables that organize our understanding. The difference between expected and observed values is constitutive for empirical testing of models in the social sciences.

**The biological perspective**

As noted, Shannon (1945) considered redundancy as "excess information" needed for error correction. Living systems need room for autonomous "error correction"—in other words, adaptation—and therefore redundancy; that is, options are needed that have not yet been realized. Brooks & Wiley (1986, p. 43) visualized the biological perspective on a system that continuously proliferates realizations, but in which order can increase "so long as the realized entropy increases at a slower rate than the maximum possible entropy" (at p. 40). In this case, the redundancy ([$H_{max} - H_{obs}$] in Figure 1) can increase even more than the realized entropy ($H_{obs}$).

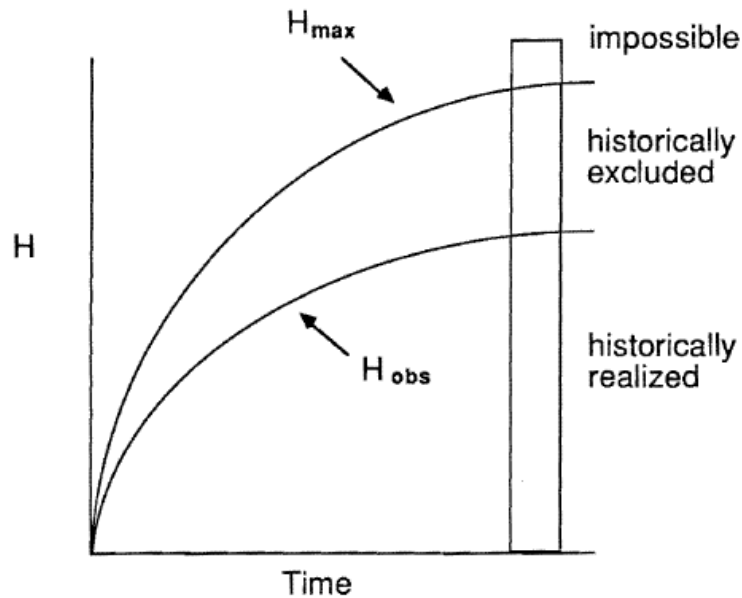

**Figure 1**: the development of information capacity, information content, and redundancy over time. Source: Brooks & Wiley, 1986, p. 43.

In the legend on the right side of this figure, "historically realized" options are first distinguished from "historically excluded" ones. System states above $H_{max}$ are labeled as "impossible." However, the authors added that "what is impossible at one time period may become possible at a later time period" (p. 43).

Kauffman (2000, at pp. 142f.) discussed the border between the historically realized and the excluded as a frontier along which "adjacent possibilities" can be realized step-by-step; in other words, along a historical trajectory. Elaborating on the example of macro-molecules, Kauffman argued that "the adjacent possible is indefinitely expandable. Once members have been realized in the current adjacent possible, a new adjacent possible, accessible from the enlarged actual that includes the novel molecules from the former adjacent possible, becomes available." In a further reflection (at p. 258) he added that "by mere constructive interference" the various trajectories may resonate into a phase transition about which "one can hope" that it provides evolutionary advantages. However, such an interference would remain coincidental: it would happen historically as an event, but the evolutionary mechanism of this change were not specified.

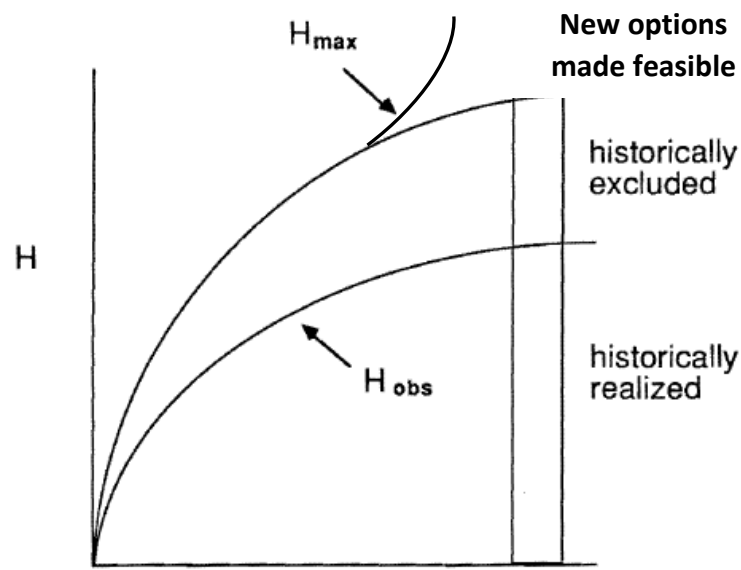

**Figure 2**: Hitherto impossible options are made possible because of cultural and technological evolution.

Using the example of transportation over the Alps, we argued above that cultural evolution may lead to changes in the number of options by orders of magnitude different from biological evolution (see Figure 2), which is historically constrained because the biological genotype is hard-wired as DNA. The cultural system, however, develops evolutionarily by refining the communication in terms of other (non-material) codes. These codes are first socially constructed, can then be generalized symbolically (Parsons, 1968), and used for a reconstruction.

The cultural level provides a regime that develops evolutionarily—above the level of phenotypical instantiations—and with a dynamics different from "natural selection." What was previously impossible can be made possible by constructing technological artifacts (e.g., airplanes) based on advances in the relevant sciences (Rosenberg, 1982). However, such a reconstruction using codes in the communication cannot be modeled using the biological model; the latter is the special case in which the codes are given.

From a biological perspective, redundancy has mainly been considered as a residual because non-information cannot be observed nor interpreted. Ulanowicz (2014) compared this refusal to specify redundancy with "negative" or "apophatic" theology:[4] the *apophasis* ($A$)—of the redundancy—cannot teach us anything about historical events, unlike the observable information which the author labels $D$ as an abbreviation of *didactic* instruction. The difference $\Phi$ between $D$ and $A$ ($\Phi = D - A$) can be considered as a measure of the degree to which the interactions (mutual informations) "remain independent of each other, i.e., the *lack* of constraints among the flows" (p. 24). However, the *apophasis* is limited by the biological constraint that $\Phi$ is positive (Ulanowicz, 1986, p. 92). A biological system with more options than realized ($A > D$) would be vulnerable to such perturbations that a catastrophe would be unavoidable (Ulanowicz, 2014, p. 26).

In our opinion, psychological and social systems are able to operate with this type of volatility on top of biological systems (such as bodies and populations) carrying the communications in physical terms. Reflexivity provides these systems with a next-order buffering capacity so that they can maintain their identities beyond historically observable stability. In psychological (action) systems stability is supported by the underlying body as a biological substrate. Inter-human communication—the next level—can contain the uncertainty, since the (latent) codes of the communication provide additional stability by structuring the communications beyond their local instantiations (Giddens, 1984).

---

[4] In apophatic theology, one can specify God only in terms of what He is not. Furthermore, ancient sages noted that the spaces between the words in the first five books of the Old Testament Bible (Torah) are as important as the words.

In summary: neither the engineering discourse of information theory nor the evolutionary discourse among biologists has specified the dynamics of redundancy other than as a residual term. However, this specification is central to understanding the dynamics of reflexivity in inter-human communications and the role of communicative agency both reflecting on and contributing to these exchanges.

**Codificiation in the vector space**

When three or more feedback arrows interact, loops are generated because of possible transitivity (Bianconi, Darst, Iacovacci, & Fortunato, 2014). Simmel (1902a and b) noted that the transition from a group of two to one of three is qualitative. In a triplet, the instantiation of one or the other relation can make a difference for the further development of the system. Furthermore, triads can be transitive or cyclic (Batagelj *et al*., 2014, pp. 53f.). Whereas transitivity can be expected to lead to organization and hierarchy, the emergence of a next-order level is based on recurrent cycles.

From a system's perspective, the network of relations—a graph—can also be considered as a vector space (Salton & McGill, 1983). The vector space reveals a topology different from the network of relations. A system of relations can already be considered as a semantic domain (Maturana, 1978, p. 49). However, semantics are not based on relations in the network space, but on relations among *patterns of relations* or, in other words, correlations. Meanings can be shared without the instantiation of a direct relation. Two synonyms, for example, can have the same position in the vector space, yet never co-occur in a sentence as a relation.

In the case of a single relation, the relational distance is not different from the correlational one; but in the case of three (or more) interacting nodes (Figure 3), distances in the vector space can be very different from distances in the network (e.g., geodesics). The structure of correlations spans the vector space as a latent background that provides meanings to the information exchanges in relations. As Giddens (1979, at p. 64) formulated: "Structures exist paradigmatically, as an absent set of differences, temporarily 'present' only in their instantiations, in the constituting moments of social systems."

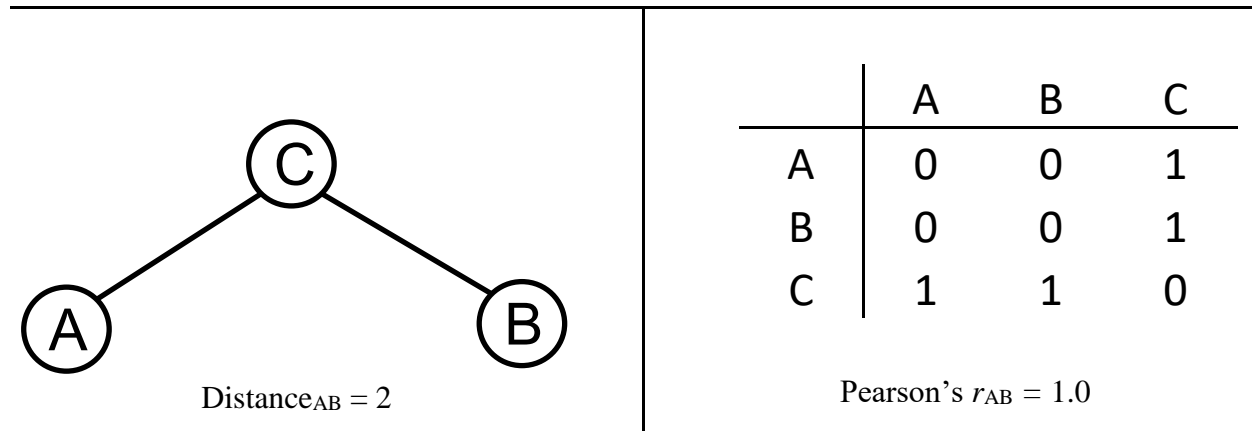

Distance$_{AB}$ = 2                       Pearson's $r_{AB}$ = 1.0

**Figure 3**: Relational distance between and structurally equivalent positions of A and B

Note that the correlations are based on the same distributions of events, but their representation in the vector space is different from a graph in the network space of observable relations. The graph in the left-hand panel of Figure 3 represents the empirically observable nodes and links—the ones—while the zeros in the right-hand panel are needed for defining the vector space. In other words, meaning is not added to the information, but the events are modeled differently and thus considered from a different perspective. As against Shannon-type information which flows linearly from the sender to the receiver, one can expect meanings to loop, and thereby to develop next-order dimensionalities (Krippendorff, 2009a and b).

What a meaning means is also contextual, and thus depends on the differentiations among these contexts. The communications can be stacked, ordered, or grouped. Whereas stacking is the default, ordering is based on transitivity, and grouping on cycles. We follow Luhmann (1995) by denoting these three forms of differentiation as segmented, stratified, and functional differentiation. According to Luhmann, segmentation prevails in primitive societies, stratification in High Culture (e.g., empires), and functional differentiation would be the hallmark of modern societies.

When the communications remain uncorrelated segments, a primitive language is naturally induced (Pinker, 1994). A High Culture requires the communication to be coded in a language which can carry symbols across border (Innis, 1950). Modern societies are based on further differentiation among the codes. The hierarchical order is then differentiated into a set of orthogonal axes, each with its own semantics.

In her book *Cultural Evolution,* Distin (2010, at p. 86) distinguishes between representations as in natural languages and meta-representations in what she calls "artefactual languages":

> A metarepresentation is a representation of another representation. Its content is that other representation, and crucially this includes information about both form and content. […] The information that evolves, is information about *how we represent*. To put this another way, once we start comparing the representational features of different languages, the two systems effectively begin to compete with each other, under representational pressure.

Aartefactual languages and meta-representations are functionally different. They enable us to accelerate the communication and thus to process more complexity. In a meta-model, information can be not only provided with meaning, but also contextualized. Meta-models enable us to learn and to optimize the relations between information processing and the processing of meaning. The further specialization of the codes can be expected to drive a cultural evolution.

**The sociological perspective**

The claim that society evolves in terms of refining codes of communication brings us to the sociological perspective. An answer to the question "what is evolving?" is crucial to the specification of an evolving system (Andersen, 1994; Hodgson & Knudsen, 2011). How is the evolution of society reflected in the sociological tradition? How can these and the above perspectives be made relevant for one another? Luhmann (1982, at p. 261) specified this long-term program of theory construction as follows:

> No matter how abstractly formulated are a general theory of systems, a general theory of evolution and a general theory of communication, all three theoretical components are necessary for the specifically sociological theory of society. They are mutually interdependent. [...]
> The decisive questions now become: How are these various theories related to one another? What unifies them? How must a theory that integrates them be constructed?

Reflecting on the relations between the state and civil society during the first half of the 19$^{th}$ century, Karl Marx ([1857] 1973, p. 265) provided perhaps the first definition of society as follows: "Society does not consist of individuals, but expresses the sum of interrelations within which these individuals stand." This definition is relational (Bhaskar, 1998, p. 207). Consciousness at positions has to be explained in terms of relations since "it is not the consciousness of men that determines their existence, but their social existence that determines their consciousness" (Marx, [1859] 1977, preface).

Max Weber, ([1904] 2015, at pp. 203 ff.) noted the historicist character of this focus on relations as the drivers of socio-cultural evolution:

> In the interest of the concrete demonstration of an ideal type or of an ideal-typical developmental sequence, one seeks to make it clear by the use of concrete illustrative material drawn from empirical-historical reality. The danger of this procedure which in itself is entirely legitimate lies in the fact that historical knowledge here appears as a servant of theory instead of the opposite role. It is a great temptation for the theorist to regard this relationship either as the normal one or, far worse, to mix theory with history and indeed to confuse them with each other.

According to Weber (1917), the "Sinn der Wertfreiheit"—the commitment to value-freeness—in the social sciences serves our ability to study values as the independent variables of social change. Weber uses the metaphor of the Greek Gods: the historical is considered as a "Kampfplatz" (battle field) of "völlig unaustragbare" (completely incompatible) values with the status of meta-representations (e.g., Weber, 1919, at pp. 608f.). Anachronistically, one could say that the values can be considered as genotypical and the historical phenomena a phenotypical. However, Weber's perspective on these values was not evolutionary.

How are values constructed and how do they evolve as social objects? Talcott Parsons (1952) noted that Freud and Durkheim independently of each other identified values as social objects that are internalized at the individual level as *super-ego* (Freud, 1930) or *conscience* (Durkheim, 1912). From this perspective, society can be considered as a reality *sui generis* (Durkheim, 1894, 1938). Parsons further added three highly relevant concepts to the analysis of this domain: (1) "double contingency" as the basic operation of the social (Parsons, 1951, pp. 91 ff.; Parsons & Shils, 1951, p. 16), (2) functional differentiation, and (3) symbolically generalized media of communication.

1. With a reference to symbolic interactionism (Mead, 1934), Parsons distinguished between a first contingency in which *Ego* and *Alter* react on each other's behaviour, and a second contingency in which *Ego* expects *Alter* to entertain expectations similar to *Ego*'s own expectations (Luhmann, 1995, pp. 103 ff.). Thus, a sociological domain of symbolic meaning processing is added to the behavioral approach which focuses on observable actions (Schutz, 1967; cf. Grathoff, 1978);
2. The social system evolves in terms of functional differentiations. As is well known, Parsons (1937) specified a four-functions paradigm of social action. However, Merton (1957) and Luhmann (1982) historicized the functions creating space for many more functions than four (Leydesdorff, Petersen, and Ivanova, 2017);
3. Because of the static character of the four-function paradigm, Parsons also needed a theory of change: change is brought about by communication. In order to appreciate the symbolic character of inter-human communication, Parsons (1963a and b) proposed symbolically generalized media of communication. In a summary for an *Encyclopedia*, Parsons (1968, p. 440) formulated the essence of this idea as follows:

> Of the many aspects of the communication process in interaction, one may be singled out for special comment, namely, the role of *generalized* media. I have already mentioned language a number of times as the prototypical medium. At the cultural level it is clearly the fundamental matrix of the whole system of media. Large-scale social systems, however, contain more specialized media (if you will, specialized "languages"), such as money, power, and influence (see Parsons 1963a; 1963b). Such media, like language, *control* behavior in the processes of interaction. They do so, however, by *symbolic* means, i.e., by presenting the actor, not with an intrinsically important object, such as a food object, but with a symbolic "representation" of such an object. […] The working of money in this regard is the best-understood example of a social system medium. […] Possession of money symbolically concretizes expectations of access to gratifying objects of utility, but money is not itself such an object.

Luhmann (1974; 1995) took the next step by coupling functional differentiation to the symbolically generalized media. These media enable us to perform at a higher level than by using only common language; the latter is considered by Luhmann (2012) as a dissemination medium since it is not function-specific. However, one should not conflate the concepts of media and the symbolic communication codes that structure them (Distin 2012, p. 94; cf. Künzler, 1987). Not media, but codes can be genotypes. The codes structure the processing of meaning and thus can be expected to operate as selection mechanisms on the redundancy generated by the signification.

**The anticipatory dynamics of meaning processing**

Meaning is accorded to historical events from the perspective of hindsight; that is, against the arrow of time. Whereas entropy increases with the arrow of time, meaning processing feeds back on the information processing by generating redundancies. Hitherto unrealized options can be entertained reflexively. The emerging codes of the communication can be considered as the axes of a component space pointing to horizons of meaning. Thus, one obtains a three-layered model: the network of observable relations, the vector space in which meanings are generated, and the component space of functionally different meta-repesentations which enable us to model the relations between the other two and thus to deconstruct, improve, and renew them.

Systems which entertain a model of themselves can be considered anticipatory (Rosen, 1985); such systems are able to explore states other than the realized ones using the model that they entertain, and thus optimize the states that are instantiated. Dubois (1998) added that a strongly anticipatory system can shape its present state on the basis of the model entertained by it; in other words, self-referentially and relatively distanced from its environment. The codes then guide the developments as (symbolically generalized) meta-representations. The following three dynamics along the time axis were distinguished by Dubois (1998; cf. Dubois & Resconi, 1992; Dubois, 2003):

1. a *recursive* dynamics in which the next state is a function of the previous one ($x_t = f(x_{t-1})$; the system develops with the arrow of time;
2. an *incursive* dynamics, in which the present state of the system is a co-determinant of its further development: $(x_t = f(x_{t-1}; x_t))$;
3. and a *hyper-incursive* system in which the present state of the system is (re-)constructed on the basis of future states: $x_t = f(x_{t+1})$.

For example, the logistic equation can be written as follows:

*recursively*: $$x_t = ax_{t-1}(1 - x_{t-1}) \qquad (2)$$
*incursively*: $$x_t = ax_{t-1}(1 - x_t) \qquad (3a)$$
or $$x_t = ax_t(1 - x_{t-1}) \qquad (3b)$$
*hyper-incursively*: $$x_t = ax_{t+1}(1 - x_{t+1}) \qquad (4)$$

Whereas Eq. 2 is the standard formula of the logistic equation—well known for its bifurcation points at $a \geq 3.0$ and increasingly chaotic development for $a \geq 3.57$—Eq. 4 models a system that is driven exclusively by expectations of future states. Such a system cannot be found in nature; it is by definition a cultural construct. For example, the rule of law on the basis of the *trias politica* is such a counterfactual—attributed as an invention to Montesquieu (1748). Since these structures of expectations operate against the arrow of time, hyper-incursions (as in Eq. 4) generate redundancy or reduction of uncertainty as opposed to the generation of uncertainty in accordance with the second law of thermodynamics which prevails in natural systems (Eq. 2).

Eqs. 3a models selection processes in the present, such as those made by embodied understanding and decision-making. For example, we build on our previous understanding, but select in the present. Analogously, a new technology builds on previous technology, but the selection term in the market $(1 - x_t)$ operates in the current situation or perhaps even on the basis of expectations $(1 - x_{t+1})$.[5] Instantiations such as the new technologies are historical and thus entrained in the entropy flow. Incursion provides the mechanism for inscribing the dynamics of cultural evolution into the historical dynamics.

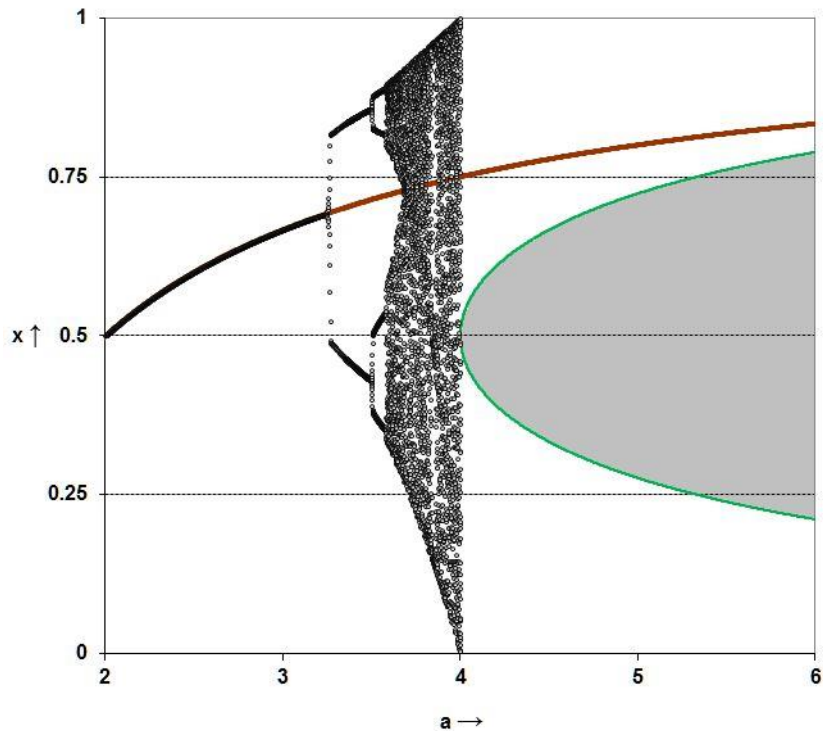

**Figure 4**: Bifurcation diagram for recursive, incursive, and hyper-incursive formulations of the logistic map; *a* is the bifurcation parameter. Source: Leydesdorff & Franse, 2009: 113.

---

[5] Advanced markets such as financial ones, for example, can be expected to operate on the basis of model-based expectations more than on the basis of previous states of the industry.

Figure 4 shows the bifurcation diagram of the logistic map using the set of Eqs. 2, 3a, and 4 for modeling (*i*) the biological domain (*a* < 4), (*ii*) the psychological one—or equivalently, another system following a trajectory in history—and (*iii*) the sociological one of structures of expectations operating for *a* > 4, respectively. On the left side one recognizes the well-known bifurcation diagram for *a* < 4; for values of *a* larger than four, this natural system vanishes in complete chaos. However, the opposite is true for the map of the cultural system on the right side (*a* > 4) based on Eq. 4.

Eq. 4 can be rewritten as follows:

$$x_t = ax_{t+1}(1 - x_{t+1}) \qquad (5)$$

$$x_t = ax_{t+1} - ax_{t+1}^2$$

$$ax_{t+1}^2 - ax_{t+1} + x_t = 0$$

$$x_{t+1}^2 - x_{t+1} + x_t/a = 0$$

In other words, Eq. 4 has two solutions for *a* > 4 which coincide for *a* = 4:

$$x_{t+1} = \tfrac{1}{2} \pm \tfrac{1}{2}\sqrt{[1 - (4/a)\, x_t]} \qquad (6)$$

With more than a single solution, an instantiation of this system requires a decision (for all values of *a* > 4). The decision makes the system historical, but requires an incursive agent who is able to perceive the options (Eq. 3a). Whereas the hyper-incursive system develops as an evolutionary regime at the next-order level generating options, the incursive system develops historically along its steady state *x* = (*a* − 1)/*a* for all values of *a*.[6] The brown line in Fig. 3

---

[6] One can rewrite Eq. 3a as follows:

$$x_{t+1} = ax_t(1 - x_{t+1}) \qquad (n1)$$
$$x_{t+1} = ax_t - ax_t x_{t+1} \qquad (n2)$$
$$x_{t+1}(1 + ax_t) = ax_t \qquad (n3)$$
$$x_{t+1} = ax_t/(1 + ax_t) \qquad (n4)$$

The steady state for $x_{t+1} = x_t$ follows:

$$1 = a/(1 + ax) \qquad (n5)$$
$$1 + ax = a \qquad (n6)$$

represents this function: the line crosses the nature/culture divide at $a = 4$ while it contains both a historically embedded component and a reflexive one. Its individual identity develops along a trajectory. The social system, however, is not an individual but can be considered as a "dividuum" (Luhmann, 1995, p. 461; cf. Nietzsche ([1878] 1967, p. 76). The system evolves in terms of distributions.

The carriers of the communication can be expected to entertain different perspectives on the social system of expectations evolving among them. The two processes of signification and codification operate as feedback mechanisms upon each other. As Langton (1989, p. 31) formulated: "local behavior supports global dynamics, which shapes local context, which affects local behavior, which supports global dynamics, and so forth." If the one feedback is dominant, the resulting process has a positive sign; but in the other case the sign is negative. The two dynamics are coupled in the incursive equation (Eq. 3a) providing a reference to both prior and expected states. The first reference generates entropy; the second reduces uncertainty. The resulting trade-off can be modeled using mutual information and mutual redundancy in interactions among systems.

**Redundancy versus information generation**

Figure 5 visualizes two sets of messages #1 and #2 with an overlap.

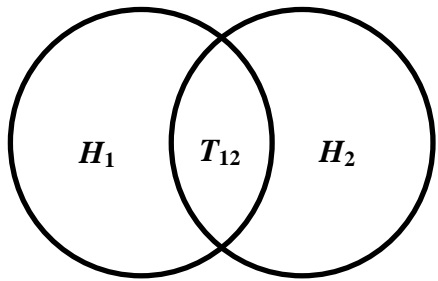

---

$$x = (a - 1)/a \qquad (n7)$$

The steady state of Eq. 3b is the same as that of Eq. 3a (Leydesdorff & Franse, 2009).

**Figure 5**: Mutual information between two sets of messages.

The sum of the expected information contents of #1 and #2 is:

$$H_{12} = H_1 + H_2 - T_{12} \tag{7}$$

In Eq. 7, one subtracts the mutual information ($T_{12}$) because otherwise this overlap would be counted twice. However, if one sums the two sets as whole sets including the overlap, one obtains $Y_{12} = H_1 + H_2 + T_{12}$. In other words, mutual redundancy has the opposite sign. $T_{12}$ is Shannon-type information and therefore positive; mutual redundancy $R_{12}(= -T_{12})$ is thus expressed in terms of negative bits of information (Yeung, 2008, pp. 59f.).

Eq. 7 can be written in a more general format:

$$T_{12} = \sum_{i=1}^{n=2} H(x_i) - H(x_1, x_2) \geq 0 \tag{8}$$

In the case of more than two dimensions, one has to correct also for the overlaps among the overlaps as follows:

$$\sum_{i=1}^{n=2} H(x_i) - H(x_1, x_2) = T_{12} \geq 0 \tag{8a}$$

$$\sum_{i=1}^{n=3} H(x_i) - H(x_1, x_2, x_3) = \sum_{ij}^{3} T_{ij} - T_{123} \geq 0 \tag{9}$$

$$\sum_{i=1}^{n=4} H(x_i) - H(x_1, x_2, x_3, x_4) = \sum_{ij}^{6} T_{ij} - \sum_{ijk}^{4} T_{ijk} + T_{1234} \geq 0 \tag{10}$$

(…)

$$\sum_{i=1}^{n} H(x_i) - H(x_1, \ldots, x_n) =$$
$$\sum_{ij}^{\binom{n}{2}} T_{ij} - \sum_{ijk}^{\binom{n}{3}} T_{ijk} + \sum_{ijkl}^{\binom{n}{4}} T_{ijkl} - \cdots + (-1)^n \sum_{ijkl\ldots(n)}^{\binom{n}{n}} T_{ijkl\ldots(n)} \geq 0 \tag{11}$$

Note that the sign of mutual information alternates with increasing dimensionality (Krippendorff, 2009b, p. 670). It follows from Eq. 11 that mutual information in *n* dimensions is:

$$\sum_{ijkl\ldots(n)}^{\binom{n}{n}} T_{ijkl\ldots(n)} = [\sum_{i=1}^{n} H(x_i) - H(x_1, \ldots, x_n)]$$

$$-[\sum_{ij}^{\binom{n}{2}} T_{ij} - \sum_{ijk}^{\binom{n}{3}} T_{ijk} + \sum_{ijkl}^{\binom{n}{4}} T_{ijkl} - \cdots + (-1)^{1+n} \sum_{ijkl\ldots(n-1)}^{\binom{n}{n-1}} T_{ijkl\ldots(n-1)}] \quad (12)$$

And mutual redundancy (with the opposite sign) is:

$$R_n = (-1)^{1+n} T_{1234\ldots n} = -[\sum_{i=1}^{n} H(x_i) - H(x_1, \ldots, x_n)]$$
$$+[\sum_{ij}^{\binom{n}{2}} T_{ij} - \sum_{ijk}^{\binom{n}{3}} T_{ijk} + \sum_{ijkl}^{\binom{n}{4}} T_{ijkl} - \cdots + (-1)^{1+n} \sum_{ijkl\ldots(n-1)}^{\binom{n}{n-1}} T_{ijkl\ldots(n-1)}] \quad (13)$$

The left-bracketed term of Eq. 13 is necessarily negative entropy (because of the subadditivity of the entropy), while mutual information in the releations contribute a second term on the right which is positive (see the set of Equations 8 to 11 above).

In other words, we model here the generation of redundancy on the one side versus the historical process of uncertainty generation on the other. The result is an empirical balance between positive and negative terms: when the resulting $R$ is negative, evolutionary dynamics prevails over historical organization, whereas a positive $R$ indicates conversely a predominance of historical organization over evolutionary self-organization. In empirical cases, one is thus able to test which combination of dimensions generates redundancy. We refer to a series of studies of national and regional systems of innovation for empirical details (e.g., Leydesdorff & Strand, 2014). However, the model is not limited to studying systems of innovation, but can be used whenever three or more independent dimensions can be specified as interacting.

**Summary and conclusions**

Unlike the DNA containing the genotype of biological evolution, the codes of cultural evolution are not given as observables, but have been constructed and are reconstructed as this evolution unfolds. The codes provide meanings to the information with reference to horizons of meaning. When the horizons of meaning are differentiated, the codes can be generalized symbolically as meta-representations organizing "artefactual languages" (Distin 2010). These media provide a functional specialization of natural languages (Parsons, 1968). Natural languages presume the development of an integrating code serving the dissemination of meaningful information (Luhmann, 2012).

Redundancy and uncertainty add up to the maximum entropy. Providing the information with meaning, however, proliferates the number of options. Codification structures the generation of meaning and thus operates as a selection mechanism. We have followed "*Darwin's Conjecture*" (Hodgson & Knudsen, 2011) of evolution theory as a general theory: phenotypical variation has to be specified differently from genotypical selection. From this perspective, biological evolution can be considered as the special case where the genotype is observable as code (DNA) and is therefore different from the variation (Künzler, 1987). Although the codes have to be constructed in communications, the cybernetic principle hypothesizes control emerging gradually as top-down feedback (Ashby, 1958; von Foerster, 1960; cf. Pickering, 2010). When this feedback is strong enough to take over control, additional redundancy can be generated by sharing meanings from different perspectives.

Luhmann (1982) conjectured that evolution theory, systems theory, and communication theory can be combined programmatically from a sociological perspective. However, he abandoned the information-theoretical perspective on the measurement. In our opinion, it is worth a try to operationalize the theoretical construct from the perspectives of evolution and communication theory and eventually to aim at measurement in information-theoretical terms. Shannon's (1948) *mathematical* theory of communication is so abstract and encompassing that its application from an engineering perspective can be considered as a special case. The engineering perspective only touches on the role of redundancy—needed for error correction—but is not able to pursue the analysis of intangibles such as meanings (Weaver, 1949). We have shown that the biological perspective also falls short when addressing this objective.

The social sciences and humanities, however, study the domains of meanings and expectations as a second contingency in interhuman communications in addition to the study of behavior in the first contingency. The distinction is analytical because the differences among the layers emerges in history. Languaging as behavior in a biological domain can not be distinguished from a semantic domain by an external observer (Maturana, 1978, p. 49). The "symbolic species" (Deacon, 1997), however, can turn from the external to the internal perspective "infra-reflexively" (Latour, 1988, at pp. 169 ff.), abstract from the information content of events, and invest symbolic value in the appreciation. When the dynamics of expectations—generating redundancies—prevail over the historical construction generating entropy, mutual redundancy is negatively signed because the relative uncertainty is reduced by increasing the redundancy.

These mechanisms were further clarified by using the theory and computation of anticipatory systems (Dubois, 1998). The feedback of meaning on information inverses the time axis leading to incursion as different from recursion. An incursive system uses its present state, for example, for selection from the perspective of hindsight. A system that uses future states for its current construction can then be considered "hyper-incursive." Hyper-incursivity cannot be natural and observable, but can only be entertained reflexively as counter-factuals which are constructed in interhuman communications. The meta-representations (Distin, 2010) or, in other words, symbolically generalized codes of communication function as "genotypes" stabilizing the volatile processing of meaning from above; that is, by controlling the coding as control mechanisms. Since these genotypes are not hard-wired, however, they can further evolve in relation to the variation provided by the signification. Instantiations require an incursive routine, including their historical reconstruction in the present.

A focus on redundancy—the "absent options" (Deakon, 2012; cf. Giddens, 1979, p. 64)—has normative implications. For example, the number of options available to an innovation system for realization may be as decisive for its future as the historically already-realized innovations. The evaluation of mutual redundancy provides us with a measure that indicates whether it is the self-organization of the meta-representations that prevails or the historical organization of the instantiations (Petersen *et al.*, 2016).

Our programmatic analysis—"Towards a Calculus of Redundancy"—is limited since bottom-up: relevant elements for developing the perspective have been brought together. However, the integration has remained intuitive. Furthermore, the focus has been on the operationalization of the duality "meaning-information" more than on the triad "knowledge-meaning-information." A top-down approach could begin with Husserl's (1929) horizons of meaning which structure the *res cogitans*—that about which we are uncertain—as distinguished by Descartes (1637) from *res extensa*—the external world. Luhmann (1986) proposed operationalizing Husserl's inter-subjective intentionality as communication in a supra-individual domain. We argued that this domain processes redundancy in a socio-cultural evolution on top and increasingly in control of the biological one.

The analytical distinctions between information, meaning and its codification, and discursive knowledge are especially pertinent for the relations between science studies and bibliometrics. Whereas the study of the sciences as cognitive constructs cannot leave the data discursively "unpacked" (Mulkay *et al*., 1983), reductionist programs can consider ("big") data as a material fundament and the distinction between *cogitans* and *extensa* as "*Descartes' Error*" (Damasio, 1994). However, the reduction of meaning processing to information—for example, in semantic maps or also using topic modeling—does not sufficiently account for the different dynamics of information and meaning processing. The sciences are not to be considered as haymaking machines which collect facts according to standardized procedures, but as developing conceptual apparatuses. Co-occurrences of words, for example, are embedded in changing (theoretical) contexts (cf. Lesk 1969; Salton 1970). A knowledge representation based on the analysis of textual (co-)occurrences is not able to account for the dynamics of the sciences in terms of structured (co-)absences at various levels of aggregation (Hesse, 1980; Leydesdorff, 1997; Liddy, 1988).

**Acknowledgement**

We are grateful to two anonymous referees for their stimulating comments.

Shannon, C. E. (1948). A Mathematical Theory of Communication. *Bell System Technical Journal, 27*, 379-423 and 623-656.
Shannon, C. E., & Weaver, W. (1949). *The Mathematical Theory of Communication*. Urbana: University of Illinois Press.
Theil, H. (1972). *Statistical Decomposition Analysis*. Amsterdam/ London: North-Holland.
Ulanowicz, R. E. (1986). *Growth and Development: Ecosystems Phenomenology*. San Jose, etc.: toExcel.
Ulanowicz, R. E. (2009). The dual nature of ecosystem dynamics. *Ecological modelling, 220*(16), 1886-1892.
Ulanowicz, R. E. (2014). Reckoning the nonexistent: Putting the science right. *Ecological modelling, 293*, 22-30.
Varela, F. J. (1979). *Principles of biological autonomy*. Amsterdam: North Holland.
von Foerster, H. (1960). On Self-organizing Systems and Their Environments. In M. C. Yovits & S. Cameron (Eds.), *Self-Organizing Systems* (pp. 31-50). London: Pergamon Press.
Von Foerster, H. (1982). *Observing Systems* (with an introduction by Francisco Varela, ed.). Seaside, CA: Intersystems Publications.
Weaver, W. (1949). Some Recent Contributions to the Mathematical Theory of Communication. In C. E. Shannon & W. Weaver (Eds.), *The Mathematical Theory of Communication* (pp. 93-117.). Urbana: University of Illinois Press.
Weber, M. (1904). Die Objektivität sozialwissenschaftlicher und sozialpolitischer Erkenntnis *Gesammelte Aufsätze zur Wissenschaftslehre* (pp. 146-214). Tübingen: Mohr, ³1968.
Weber, M. (1917). Der Sinn der 'Wertfreiheit' der soziologischen und ökonomischen Wissenschaften. *Gesammelte Aufsätze zur Wissenschaftslehre* (pp. 489-540). Tübingen: Mohr, ³1968.
Yeung, R. W. (2008). *Information Theory and Network Coding*. New York, NY: Springer.